\documentclass[twocolumn]{aastex62}
\usepackage{ragged2e}
\usepackage{natbib}

\newcommand{\corr}{\textcolor{black}}
\newcommand{\dorr}{\textcolor{black}}

\shorttitle{Stealth CMEs from active regions}
\shortauthors{O'Kane et al.}
\begin{document}

%\title{Understanding the origins of stealth coronal mass ejections}
\title{Stealth Coronal Mass Ejections from Active Regions}

\author[0000-0002-8806-5591]{Jennifer O'Kane}
\affil{Mullard Space Science Laboratory, UCL, Holmbury St Mary, Dorking, Surrey, RH5 6NT, UK}

\author[0000-0002-0053-4876]{Lucie Green}
\affil{Mullard Space Science Laboratory, UCL, Holmbury St Mary, Dorking, Surrey, RH5 6NT, UK}

\author[0000-0003-3137-0277]{David M.~Long}
\affil{Mullard Space Science Laboratory, UCL, Holmbury St Mary, Dorking, Surrey, RH5 6NT, UK}

\author[0000-0002-6287-3494]{Hamish Reid}
\affil{SUPA School of Physics and Astronomy, University of Glasgow, Glasgow, G12 8QQ, UK}
\nocollaboration



\begin{abstract}

Stealth coronal mass ejections (CMEs) are eruptions from the Sun that have no obvious low coronal signature. These CMEs are characteristically slower events, but can still be geoeffective and affect space weather at Earth. Therefore, understanding the science underpinning these eruptions will greatly improve our ability to detect and, eventually, forecast them. We present a study of two stealth CMEs analysed using advanced image processing techniques that reveal their faint signatures in observations from the extreme ultraviolet (EUV) imagers onboard the Solar and Heliospheric Observatory (SOHO), Solar Dynamics Observatory (SDO), and Solar Terrestrial Relations Observatory (STEREO) spacecraft. The different viewpoints given by these spacecraft provide the opportunity to study each eruption from above and the side contemporaneously. For each event, EUV and magnetogram observations were combined to reveal the coronal structure that erupted. For one event, the observations indicate the presence of a magnetic flux rope before the CME's fast rise phase. We found that both events originated in active regions and are likely to be sympathetic CMEs triggered by a nearby eruption. We discuss the physical processes that occurred in the time leading up to the onset of each stealth CME and conclude that these eruptions are part of the low-energy and velocity tail of a distribution of CME events, and are not a distinct phenomenon.

\end{abstract}



\section{Introduction} \label{sec:intro}

Coronal mass ejections (CMEs) are large eruptions of solar plasma, embedded with the solar magnetic field. Upon occurrence, typically one or more signatures are observed in the lower solar atmosphere, such as filament eruptions, solar flares, post-eruptive arcades, EUV dimmings and EUV waves that enable the CME source region to be identified \citep[see][for an overview]{webb2012coronal}. Observations from the Earth's viewpoint means that a white light CME with no observable low coronal signatures is normally assumed to be a backsided event. However, the launch of the twin Solar Terrestrial Relations Observatory (STEREO) spacecraft \citep{kaiser2008stereo} enabled the CME propagation direction to be determined, and therefore the identification of which side of the Sun the CME originated from, using geometric triangulation techniques.
%CMEs with no such signatures to be classed as a front-sided or back-sided event using geometric triangulation techniques. 
\citet{robbrecht2009no} named those eruptions that are seen in coronagraph data but which leave no observable signatures in the low corona as ``stealth" CMEs.

Stealth CMEs typically have plane-of-sky speeds less than 500~km~s$^{-1}$ \citep{d2014observational} and are frequently found to originate from quiet Sun regions \citep{ma2010statistical} and regions close to open magnetic field \citep{nitta2017earth}. They are fairly common; for example, \citet{ma2010statistical} found that 1/3 of Earth-sided CMEs at solar minimum had no distinct signatures, while the statistical study of \citet{wang2011statistical} during the solar minimum of 1997-1998 found that 16\% of  front sided CMEs showed no signatures of the eruption on disk.  Additionally, \citet{kilpua2014solar}, in a study of 16 interplanetary CMEs (ICMEs) from 2009, found that 10 ICMEs were stealth events. Despite their slow speeds, stealth CMEs can be the source of geomagnetic activity. \citet{zhang2007solar} studied 77 geomagnetic storms associated with ICMEs that occurred in solar cycle 23. Of these events, nine could not be associated with phenomena occurring on the solar disk. A more recent study by \citet{nitta2017earth} focused on a set of stealth events that caused disturbances at 1AU, three of which produced Dst (Disturbance storm time) values greater than -100nT, indicative of a moderate geomagnetic storm.  

 The geomagnetic impact and frequency of stealth CMEs means that interest in this type of eruption is growing. There are many open issues, including the fundamental question of whether or not stealth CMEs are different from non-stealth events. A review by \citet{howard2013stealth} led the authors to propose that stealth CMEs sit at the lower energy end of a continuous spectrum of events, and originate from streamer blowouts. The authors suggest that the lack of observable signatures is likely due to instrumentation limitations and that the classification of this type of event is a purely observational one. If this is the case, then the trigger and driver mechanisms that are currently proposed for CMEs should be relevant and, once the eruption is underway, observational signatures of the CSHKP standard flare model \citep{carmichael1964process,sturrock1966model,hirayama1974theoretical,kopp1976magnetic} should be sought with appropriate instrumentation or image processing techniques.

 Theories and models of CMEs focus on key aspects such as the specific magnetic configuration of the non-potential pre-eruptive field (that stores the free magnetic energy used to power the eruption), its evolution due to photospheric flows and/or flux emergence (the energy storage phase), and whether ideal or non-ideal processes are able to affect the stability of the field and bring it to the point of eruption (the energy release phase). Models include the breakout model  \citep{antiochos1999model} in which strong shear is invoked within the central arcade of a multipolar system. The shear leads to inflation of the core field followed by reconnection with the overlying arcade. This removes overlying field and allows the sheared arcade to erupt if a second phase of reconnection occurs within the sheared arcade, transforming it into a flux rope. The tether cutting model also involves a sheared arcade in which runaway reconnection both builds and ejects the flux rope \citep{moore2001onset}. On the other hand, some models require the pre-eruption field to be that of a magnetic flux rope. The flux rope can become unstable due to the torus instability if the gradient of the field overlying the curved flux rope falls sufficiently rapidly with height \citep{kliem2006torus}. Removal of overlying field, resulting from a nearby CME as in the sympathetic eruption model \citep{torok2011}, may create this condition, as could an increase in flux of the rope which would raise the structure. A comprehensive review of CME models and \corr{their observational indicators may} be found in Table 1 in \citet{green2018origin}.

 The challenge then for stealth CME  studies is to try and determine whether existing data can be used to investigate the processes involved, and whether aspects of the CME models discussed above are operating. A variety of observational signatures can be utilised to do this. For example, photospheric magnetic field can be used to identify both sustained shear flows along polarity inversion lines, and flux emergence. The configuration and evolution of the coronal field can be studied using emission structures observed in EUV or X-ray data. This may include for example the identification of a pre-existing flux rope in the corona before CME onset through EUV or soft X-ray sigmoidal structures \citep{green2009,green2011photospheric,james2017disc}, and hot flux ropes \citep{cheng2011observing,zhang2012observation,patsourakos2013direct,nindos2015common}. In all these cases, the observed flux rope forms via reconnection either at a low altitude as in the \citet{van1989formation} model or higher up in the corona \citep{james2017disc,james2018} with corresponding observational signatures.  
However, such investigations are predicated on the identification of the correct source region for the stealth CME.

%\startlongtable
\begin{deluxetable*}{cccccc}[!ht] \label{table:1}
\centering
\tablecaption{Details of the instruments used within this study. SE1 = Stealth event 1, 27-Oct-2009. SE2 = Stealth event 2, 03-Mar-2011.}
\tablehead{\colhead{Instrument/Spacecraft} & \colhead{Type} & \colhead{Used for event} & \colhead{Wavelengths}  & \colhead{Resolution}  & \colhead{FOV}  \\ \colhead{} & \colhead{} & \colhead{} & \colhead{(\AA)} & \colhead{(arcsec)} & \colhead{(R$_{\odot}$)}}
\startdata
EIT/SOHO & EUV imager & SE1 & 195 & 5.2  & 0-1.5 \\
MDI/SOHO & Magnetogram & SE1 & N/A & 4  & 0-1 \\
LASCO C2/SOHO & WL coronagraph & SE2 & N/A & 47 & 1.5-6 \\
EUVI/STEREO & EUV imager & SE1, SE2 & 195, 304 & 3.2  & 0-1.7  \\
COR1/STEREO & WL coronagraph & SE1, SE2 & N/A & 15  & 1.5-4   \\
AIA/SDO & EUV imager & SE2 & 94, 131, 171, 193, 211, 304 & 1.2 & 0-1.5  \\
HMI/SDO & Magnetogram & SE2 & N/A & 1 & 0-1  \\
Nancay Radioheliograph & Radio Interferometer  & SE2 & 150 MHz & 200 & 0-2 \\
\enddata
\end{deluxetable*}

 Whilst there are many ways in which a CME can be formed, once the eruption is underway there is consensus within the community as to how the magnetic field evolves. The erupting structure proceeds through a sequence of distinct evolutionary phases in its kinematics. First, a slow rise of around 10 kms$^{-1}$, possibly where the stability of the field is lost, second, a rapid acceleration up to velocities of 100s km$^{-1}$ to 1000s km$^{-1}$ when the main energy release (and magnetic reconnection) occurs, and third, propagation into the heliosphere \citep{zhang2001temporal,zhang2006,vrsnak2008}. Any soft X-ray flare emission rises sharply during phase 2, indicating the close coupling between the flare reconnection and CME acceleration. Indeed, phase two can exhibit a variety of observational signatures, which are collectively described by the CSHKP standard model. These signatures include flare reconnection under the erupting structure that produces a post-eruption (flare) arcade and, as the core field expands, the reduction of plasma density in the lower corona (which produces dimming regions). Once the eruption is underway, so by phase two, all CME models discussed above find that the magnetic configuration is that of a flux rope regardless of the pre-CME field details. Observations of CMEs studied using coronagraph data \citep{vourlidas2013} and in situ \citep{burlaga1981magnetic} indeed find flux ropes in many cases. If stealth CMEs do not differ from other CMEs, it can be expected that they would follow such evolutionary stages, but without obvious flare emission due to the low energy release, and exhibit a flux rope configuration as they leave the Sun.

There is yet to be a clear definition of stealth CMEs with some works stating that a stealth CME is one with no low coronal signatures whilst other works define a stealth CME as one with no obvious, or very weak, low coronal signatures. Although the differences between these two classifications may at first seem trivial, stating that there are no signatures at all suggests that the signatures simply do not exist. On the other hand, if they are events with very weak and/or no obvious signatures, the events may not necessarily be fundamentally different to other CMEs, and work towards producing tools and techniques that reveal these weaker signatures can progress. A comprehensive study by \citet{alzate2017identification} showed,\corr{ using} advanced imaging processing techniques applied to coronal observations, that all 40 stealth CMEs in a catalogue developed by \citet{d2014observational} did indeed manifest themselves with one or more lower coronal signatures. This suggests that  the source regions of stealth CMEs can be found and studied.

%stealth CMEs are no different to other CMEs in terms of the physical processes involved. 

%We are going to look at our observations and compare them to what you would expect to observe in each of these cases, and compare them to observations gained from the standard model. 

 The aim of this study is to combine knowledge of the observational signatures of CMEs related to the formation and eruption of non-potential fields with the latest image processing techniques to extend the study of \cite{alzate2017identification} for two stealth CME events. Stealth event 1 occurred on 27 October 2009 and stealth event 2 occurred on 03 March 2011. Both events have had their approximate source region determined using triangulation by \cite{kilpua2014solar} and \cite{pevtsov2011coronal}, respectively. We aim to investigate whether these stealth CMEs show signatures of the CSHKP standard model, albeit faint, in order to identify the exact source region. We then look for signatures of the processes that could account for the formation and de-stabilization of the eruptive structures. In Section \ref{sec:data_analysis} we describe data used and our analysis techniques. Section \ref{sec:results} displays our findings for the two stealth CME events, and these are discussed in Section \ref{sec:discussion}. Conclusions are presented in Section \ref{sec:conclusion}.

%Can we use the standard model to explain stealth CMEs? What are the formation and initiation mechanisms of the CME? / When does the flux rope form? 

\section{Observations and Methods}
\label{sec:data_analysis}
This work uses data from the the Atmospheric Imaging Assembly \citep[AIA;][]{lemen2012} onboard the Solar Dynamics Observatory \citep[SDO;][]{pesnell2011solar}, the Extreme Ultraviolet \corr{Imager} \citep[EUVI;][]{howard2008sun} onboard the Solar Terrestrial Relations Observatory \citep[STEREO;][]{kaiser2008stereo} and the Extreme ultraviolet Imaging Telescope \citep[EIT;][]{delaboudiniere1995eit} onboard the Solar and Heliospheric Observatory \citep[SOHO;][]{domingo1995soho}. Photospheric line of sight magnetograms are obtained from the Heliseismic and Magnetic Imager \citep[HMI;][]{scherrer2012helioseismic} and the Michelson Doppler Imager \citep[MDI;][]{scherrer1995solar} onboard SDO and SOHO respectively. The CMEs are identified using data from the white-light coronagraphs, The Large Angle and Spectrometric Coronagraph \citep[LASCO;][]{brueckner1995large} and COR1/COR2 \citep[\corr{part of the SECCHI instrument suite};][]{howard2008sun} onboard SOHO and STEREO respectively. The details of instruments used in each stealth CME event are outlined in Table \ref{table:1}.

%Ref edit : Swapped the first two sentences so that when they are discussed in more detail next, it follows the same order
\corr{The graduated cylindrical shell model and stack plots of EUVI and COR1 were used to trace the CME back to the start of the eruption}. In order to increase the likelihood of being able to identify lower coronal signatures of the stealth CMEs in EUV data, the Multi-scale Gaussian Normalization technique \citep[MGN;][]{morgan2014multi} was applied and running difference images were created for both events and examined for dimming regions. \corr{The radio data was imaged using the Nan\c{c}ay Radioheliograph (NRH) \citep{kerdraon1997nanccay}}.

\subsection{Graduated Cylindrical Shell (GCS) Model}
The graduated cylindrical shell (GCS) model is an empirical model developed by \citet{thernisien2006modeling}. It is used to study the 3-D morphology, position, and kinematics of flux-rope CMEs. Flux-rope CMEs typically have a 3-part structure; a bright leading front, a dark cavity, and a bright core, where the dark cavity is representive of a flux rope within the CME structure. The geometry of the model includes cone shaped legs, a pseudo-circular front, and a circular cross section, fitted by eye using at least two different vantage points and six parameters; longitude, latitude, height, tilt angle, half angle, and ratio. The resulting GCS model is a shape similar to that of a hollow croissant, which expands in a self-similar way. The model requires at least two coronagraph images from two different spacecraft (e.g COR2-A and COR2-B) taken at the same time. 

\dorr{The GCS model used COR1 and COR2 data to find approximate source region for each stealth CME. A radial path away from the Sun is assumed, however it is noted that many CMEs in these lower regions are non-radial with respect to their source regions \citep{cremades2004three,cremades2006properties}}.
 
Typically when fitting the GCS model to the coronagraph images, one would also ensure that the \corr{flux rope} footpoints from the model match up with those determined observationally in the EUV images. Due to stealth CMEs having no obvious low coronal signatures, it is an opportunity to obtain an approximation of the location of the CME footpoints, and thus the source region of the CME. However due to both limitations with the model and possible alterations in CME direction after eruption, the source region obtained from the GCS model may not be exactly correct. Therefore it is necessary to also search for observational signatures of the eruption.

\begin{figure*}[t!]%
    \centering
    {{\includegraphics[width=8.5cm]{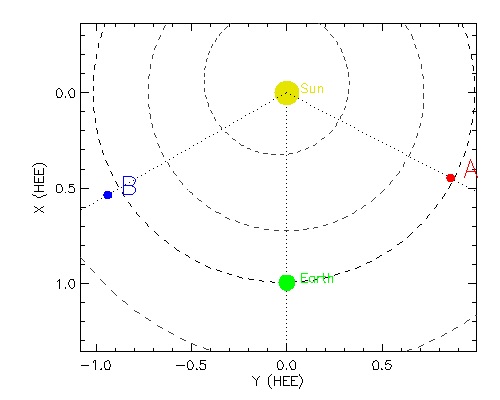} }}%
    \qquad
    {{\includegraphics[width=8.5cm]{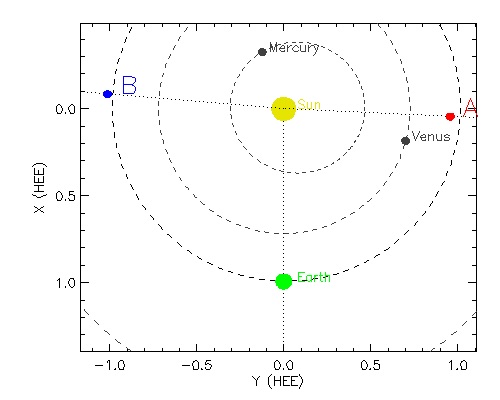} }}%
    \caption{Positions of STEREO-A and -B during on 27-Oct-2009 (left) and 03-Mar-2011 (right)}%
    \label{fig:pos}%
\end{figure*}

\subsection{Observations}
%Ref edit : This is a bit of a word vomit and will need clearing up
\corr{Stack plots of COR1 and EUVI were created for both events. The slices generated from the stack plots were radially outwards from the Sun. In each case, stack plots were created for all angles that crossed the CME structure at a variety of angular widths, as observed in COR1 field of views. All stack plots were examined for potential activity. The stack plots presented in this paper, are the radial slices that intersect through the centre of the concave-up structure, assumed to be the flux-rope cavity, as this showed the most clear propagating CME structure. } 

%Intro paragraph on difference imaging to clarify why we picked 30 minutes etc and why we used ratio images too
\corr{For each instrument, a variety of difference imaging techniques were employed at various temporal separations. The running difference and running ratio images gave the best result for our work. The running difference subtracts a following image from a leading image, whilst the running \dorr{ratio} divides a leading image from a following image. Temporal separations varied between 2 minutes and 3 hours. \dorr{Thirty} minutes proved to provide a clearer image, for capturing the dynamic motions, without having too much effect on the ambient background, whilst 3 hours proved best for capturing faint EUV dimmings.}
The first event studied here occurred on October 27th 2009, and was observed by SOHO and both STEREO spacecraft. The separation angle between STEREO-A and -B was 123$^\circ$, with STEREO-B 60$^\circ$ behind Earth and STEREO-A 63$^\circ$ ahead of Earth (left panel of Figure \ref{fig:pos}). For this event, one of the EIT wavebands was used, and two of the EUVI wavebands were used (see Table \ref{table:1}). For EIT we used a 12 minute temporal separation, and for each of the EUVI wavelengths we used 5 minute, 10 minute and 30 minute temporal separations. The second event studied here occurred on March 3rd 2011, and was observed on-disk by SDO and at the limb by both STEREO spacecraft. At this time the separation angle between STEREO-A and -B was 178$^\circ$, with STEREO-B 95$^\circ$ behind Earth and STEREO-A 87$^\circ$ ahead of Earth (right panel of Figure \ref{fig:pos}). Running time difference images were created for six of the AIA wavebands, and for two of the EUVI wavebands, as outlined in Table \ref{table:1}. For each of the AIA passbands we used 2 minute, 5 minute and 10 minute temporal separations. For each of the EUVI passbands we used 5 minute, 10 minute and 30 minute temporal separations. Longer temporal separations were necessary in order to observe dynamic structure of the stealth CMEs, which evolve at a relatively slow rate. Temporal separations were chosen based on the cadence of each instrument. The temporal evolution of the CME as it propagates outward from the Sun was tracked using a stack plot which combined the fields of view of both EUVI and COR1 from the STEREO spacecraft.
%Height-time profiles for both stealth events were created using EUVI and COR1 of the STEREO spacecraft. The height-time profiles allowed us to track the CME as it propagated outwards from the Sun. The plots were constructed using a segment with a variable angular width extending from the solar surface to the edge of the field of view. Once the height-time profile was plotted, exponential and quadratic curves were fitted to the CME curve observed in the COR1 height-time profile. The curve that fits best to the CME can tell us something about the most likely initiation mechanism of the two events. Typically, a quadratic rise profile is often  associated with a breakout model scenario, whereas the exponential rise profile is associated with an instability dominating the eruption. 

The MGN technique was applied to each of the EUV passbands listed in Table \ref{table:1}. This technique reveals faint structure in the low corona that is usually hidden as a result of bright regions that dominate over regions of the Sun with lower EUV emission. The ability to observe this fine structure is obtained by normalizing images at multiple spatial scales, using the local mean and standard deviation, and producing a weighted combination of the normalised components. The method produces detailed images similar to wavelet based techniques \citep{stenborg2003wavelet,stenborg2008fresh}, and the noise adaptive fuzzy equalization technique \citep{druckmuller2013noise}, however MGN is much more computationally efficient, faster by at least an order of magnitude, and does not require a high-performance computer. The detailed images produced by this technique can reveal fine structural changes in the low corona that are related to the formation and later eruption of the stealth CME. 

%Two faint EUV dimmings were observed in the running difference images of AIA 193\AA\ and 211\AA\ during stealth event 1. 
 Running ratio images with 3 hour temporal separations were used, in order to \corr{identify, track,} and enhance any EUV dimmings associated with the stealth CMEs. Edges of  dimmings were manually selected every half hour, and the contours of the dimming regions at each time are plotted.

%over the time period the dimmings were observed, 00:00 UT to 09:30 UT.

MDI and HMI were used to observe the evolution of the photsopheric magnetic field for the October 27th 2009 and March 3rd 2011 stealth CMEs respectively. We searched for changes in the magnetic field such as flux emergence, flux cancellation, and shearing motions that may play an important roll in the formation and initiation of CMEs, as well as observing the configuration of the magnetic field at the time of eruption. 

\corr{The NRH was used to analyse the radio emission that arose during the March 3rd 2011 stealth event.  The radio images were made using the NRH clean algorithm from data with a 1 second cadence.  The radio flux was calculated from the images using a box of length 600 arcsecs.} 

\section{Results} \label{sec:results}

\subsection{Stealth event 1 : 27-October-2009}
On 27 October 2009 the STEREO coronagraphs observed a CME that was found by \citet{kilpua2014solar} to be Earth directed, but without any low coronal signatures. The authors reported that the CME first appeared in STEREO-B (STEREO-A) COR1 at 10:30 UT (15:30 UT) on 27 October 2009, giving an estimated eruption onset time of $\sim$06:00 UT 27-Oct-2009. The CME had a plane-of-sky speed of 208 kms$^{-1}$ as seen by STEREO-A\footnote{http://solar.jhuapl.edu/Data-Products/COR-CME-Catalog.php}. \citet{kilpua2014solar} performed a multi-spacecraft forward-modelling analysis using the GCS model and applied a triangulation technique, approximating the source region of the event to be located at N03W06, and N03W10 respectively from the two methods. Although eruptive signatures were observed on the solar disk, the authors concluded that this activity was not co-spatial with the approximated source region of the stealth CME. Instead, the estimated source region is approximately half way between two active regions, one in its emergence phase (AR 11029) and one in its decay phase with no NOAA active region number assigned. 

%We re-analyse the EUV data using the MGN technique and discuss below why we propose active region 11029 to be the source region of the stealth CME.

  \begin{figure*}[t!]
\centering
    \includegraphics[width=1\linewidth]{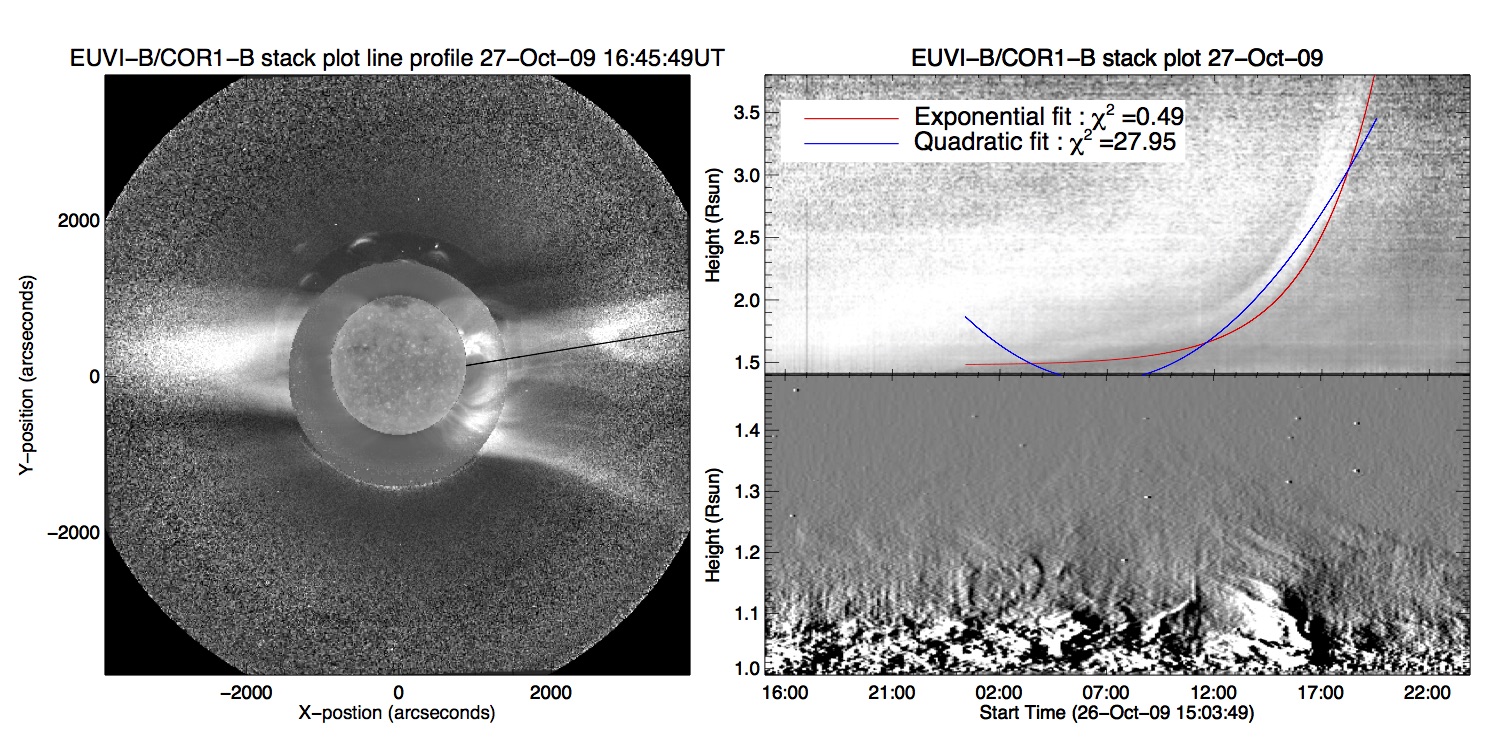}
    \caption{(Left) \corr{EUVI}/COR1 image indicating the slice used to create the stack plot (solid line). (Right) EUVI-B 195\AA~(lower) and COR1-B (upper) stack plots.  The height-time profile of the stealth CME observed on 27 October 2009 can be determined from the COR1-B data which show the underside (concave-up structure) of the stealth CME. Exponential (red line) and quadratic (blue line) fits have been applied to the COR1-B data.}
    \label{fig:ht_se2}
  \end{figure*}

  \begin{figure*}[t!]
\centering
    \includegraphics[width=1\linewidth]{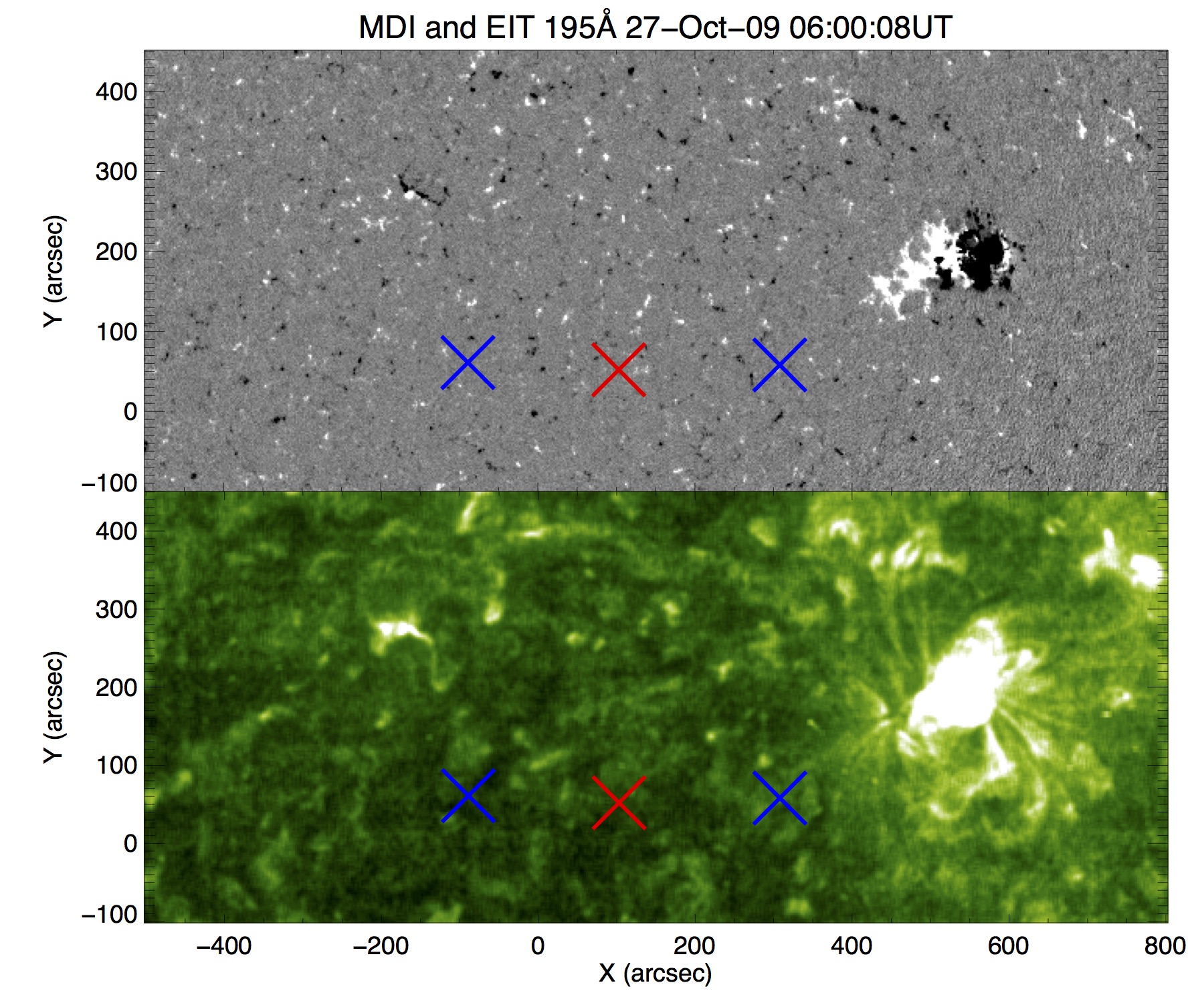}
    \caption{Top panel: MDI magnetograph image. Bottom panel: MGN processed EIT 195\AA\ image. The triangulated source region determined by \cite{kilpua2014solar} is illustrated by the red X. The footpoints of the eruption approximated by the GCS model are illustrated by the blue X's. The estimated source region is located in a region of highly \corr{dispersed} magnetic field, with no polarity inversion line, suggesting that an eruption is unlikely to have occurred here. A small decayed active region is located north-east, and a newly emerged active region (NOAA 11029) is located north-west of the estimated source region. }
    \label{fig:eit}
  \end{figure*}

\subsubsection{Analysis} %Note this heading is temporary until I think of a better one
%Stack plots --> GCS finding footpoints --> EUV observations --> HMI observations

%Stack plots
Stack plots were created to track the CME through the coronagraph and EUV data, to the surface of the Sun. Figure \ref{fig:ht_se2} shows the stack plot created using a slice of the data at an angle of 80$^\circ$ from solar north in the clockwise direction. The propagation of the concave-up section of the CME is clearly visible in the COR1-A stack plot created from a time series of these data slices, where the CME appears to exhibit a slow rise phase that is in progress by 23:00 UT 26-Oct-2009, followed by a phase of rapid acceleration $\sim$13:00 UT 27-Oct-2009. However, the EUVI-A stack plot does not show any clear upward propagating structure.
%REF EDIT : Trying to be more clear about concluding a high altitude initiation
\corr{The result that it was unable to be picked up in the field of view captured by EUVI-A, combined with the observation of the slow rise phase in the COR1 field of view, suggests that the eruption was initiated from higher altitudes.} The EUVI stack plot does however show a large brightening around the time of the rapid acceleration phase beginning, likely to be post-eruption loops associated with the stealth CME. We then fitted exponential and quadratic curves to the CME position in COR1 field of view. The curves can give indications of what mechanisms are driving the eruption \citep{d2014observational,schrijver2008observations}. Numerical simulations matched with observations have demonstrated that an exponential rise profile occurs when an instability is dominating the eruption, such as the torus or kink instability, whilst a quadratic rise profile occurs during a breakout model scenario. For this event the exponential curve produced the best fit, suggesting an instability dominating the eruption. Both the torus and kink instability require a flux rope prior to eruption, and given that the concave-up structure could be tracked for a period of hours during the slow rise phase, it is likely that in addition to the structure being initiated at a high altitude, the flux rope was also formed at a high altitude in the corona. 
%This suggests that the loops observed to form in EUV data around this time are likely to be post-eruption loops associated with the stealth CME. There appears to be an indication of the post-eruption loop formation in the EUVI data in the lower right planel of Figure  \ref{fig:ht_se2}.   %eruption was initiated several hours earlier than that approximated by \citet{kilpua2014solar}.

%Coronagraph Observations 
The stealth CME is not observed in the COR1-A images until $\sim$15:30UT on 27-Oct-2009, several hours after the eruption onset time as determined by \citet{kilpua2014solar}. The CME is also very faint and barely visible in the images, making it difficult to pinpoint the exact time that it enters the field of view. It also emerges from a bright streamer that acts to mask the CME structure. However, the stealth CME is seen more clearly, and earlier, in COR1-B, which could suggest that the eruption is closer to the limb of STEREO-B than STEREO-A. The estimated source region from triangulation techniques is of similar distance to the nearest limbs in both spacecraft, whilst the decayed active region is directly on the limb in STEREO-A and on disk in STEREO-B, and active region 11029 is directly on the limb of STEREO-B and on disk in STEREO-A.

\begin{figure*}[t!]
\centering
    \includegraphics[width=1\linewidth]{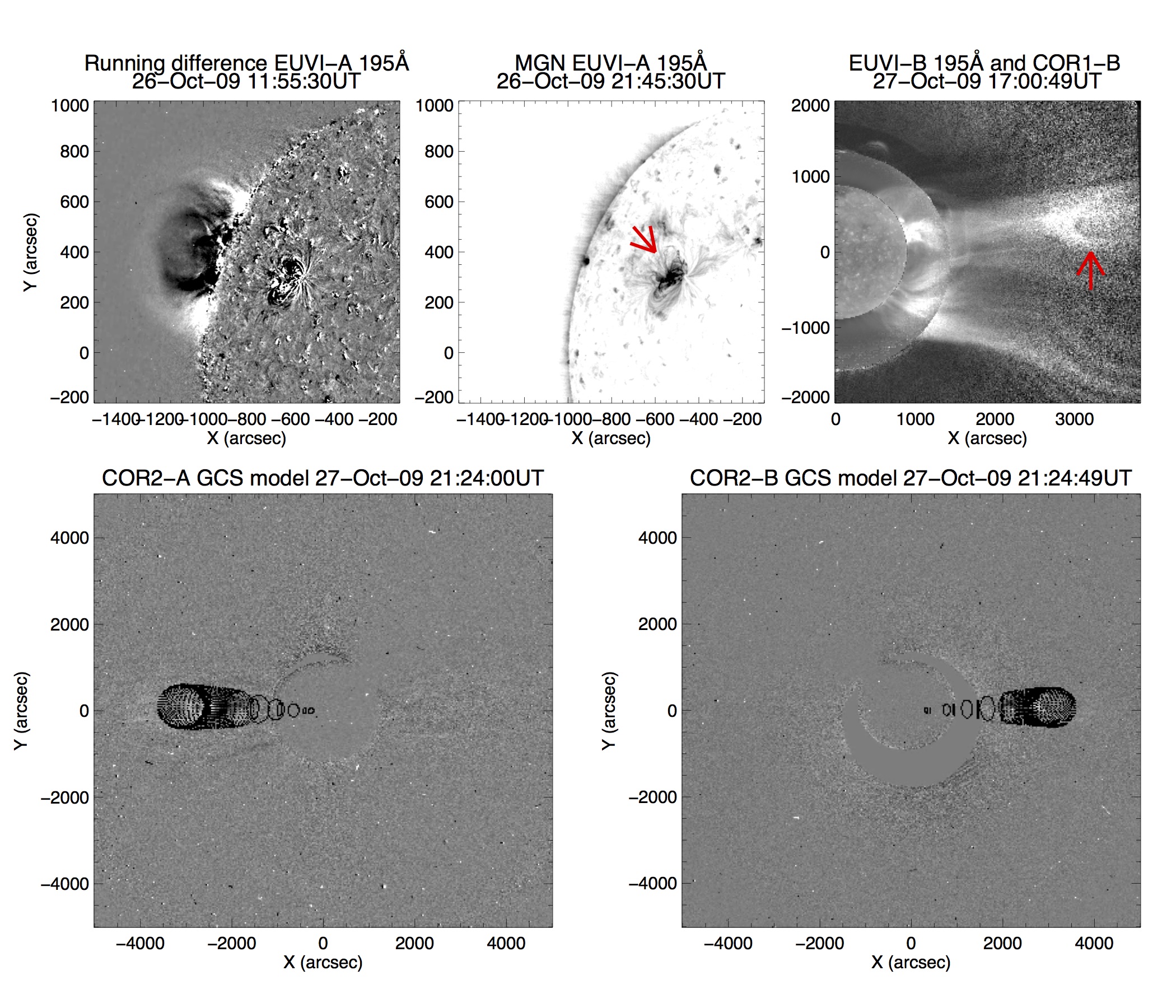}
    \caption{Top : (Left) EUVI-A 195\AA\ running difference image with a 5 minute temporal separation, showing an eruption from the decayed active region. The eruption occurs several hours before the eruption of the stealth CME. (Center) MGN processed EUVI-A 195\AA\ showing a small eruptive burst of plasma from the newly emerged active region, believed to be the source of the stealth CME. (Right) Combined EUVI-B 195\AA\ and COR1-B image. The CME is indicated by the arrow and has a dark cavity, indicative of a flux rope present in the CME. Bottom : The wire frame fitted to the CME in COR2-A (Left) and COR2- B (Right).}
    \label{fig:sta}
  \end{figure*}

%GCS MODEL  --- Something doesnt feel right having the footpoints in one figure, and the model fit in another but describng them in the same paragraph before even talking about the EUV obs....But I think the GCS should be discussed before the EUV, as its about finding an approximate source region to look into... hmmmm... 
Using COR1 and COR2 from the twin STEREO spacecraft, the wireframe of the GCS model (representing the flux rope), was fitted to the concave-up cavity structure in the coronagraph observations (Figure \ref{fig:sta}). The footpoints of the erupting structure were found to be N03E05 and N03W18 from the GCS model, illustrated by the blue X's in Figure \ref{fig:eit}, centered around the region approximated by \citet{kilpua2014solar} (red X in Figure \ref{fig:eit}) using the triangulation and the GCS model. This approximated source region is located in a quiet Sun region in the northern hemisphere. On the east and west sides of the approximated source region were a decayed active region and NOAA active region 11029, respectively (Figure \ref{fig:eit}).

%EUV observations
%Both regions exhibited multiple brightenings and eruptions and we now investigate whether any of these activity events could be signatures of the stealth CME. 
We then applied the MGN technique to the EUV data, and additionally produced running difference images using the EUV data, enabling a variety of dynamical structures to be observed in the lead up to, and following the stealth CME. The activity in the decayed active region can be summarised as: a large and clear eruption beginning $\sim$11:25 UT 26-Oct-2009 as seen in STEREO-A EUVI data  with an associated white light CME observed in COR1-A $\sim$12:00 UT (Figure \ref{fig:sta}, first panel); a very faint structure that moves outwards through the field of view of STEREO-A COR1 $\sim$21:00 UT 26-Oct-2009, this may be a part of the previous event or a separate eruption that closely follows the former; an extremely faint rising loop $\sim$07:10 UT that cannot be followed to the edge of the field of view of EUVI, since no associated post-eruption loops were observed, it is deemed to be a failed eruption. None of these activity events can be shown to be associated with the stealth CME and therefore the decayed active region is not deemed to be its source region. The activity in active region 11029 in the $\sim$ 1.5 days before the stealth CME was first observed in coronagraph data can be summarised as: small burst of bright plasma in the north of the active region $\sim$21:45 UT 26-Oct-2009; expanding loops begin forming early on 27-Oct-2009 to $\sim$05:10 UT; multiple C-class flares between 18:38 UT on 26-Oct-2009, and 11:07 UT on 27-Oct-2009 (Figure \ref{fig:sta}, second panel). During this time, active region 11029 continues to brighten and displays an ongoing reconfiguration of the loops, which may be associated with the ongoing flux emergence in the region. There is a weak dimming region to the north of the active region and on its western side seen in EIT data that begins $\sim$06:30 UT 27-Oct-2009, however neither running difference nor running ratio images could enhance this to a trackable feature. By $\sim$12:55 UT active region 11029 continues to brighten with new loops forming. From the location of active region 11029 along with the observed dimming and reconfigured field we conclude that this is the most likely source region of the stealth CME on 27 October 2009.

% note from Lucie: I am not convinced that there was a small eruption from the decayed region at $\sim$02:15 UT and $\sim$23:30 UT on 26-Oct-2009  or $\sim$01:30 UT on 27th October - so I removed them. from the list.

% A small eruption occurred in the decayed active region $\sim$02:15 UT 26-Oct-2009. The loops in AR 11029 brighten for a few hours $\sim$03:45 UT 26-Oct-2009, and the region appears to be very dynamic. A large eruption is produced from the decayed active region $\sim$11:25 UT 26-Oct-2009 and is observed in COR1-A $\sim$12:00 UT. The CME becomes faint as it propagates through the field of view of COR1-A, and it is unclear where the CME ends. There is a very faint structure that moves outwards through the field of view $\sim$21:00 UT, it is unclear if it is a separate eruption, or if it is material still being pulled out by the previous eruption. As the large eruption erupts and expands, it looks like the CME splits slightly, and the lower half could be this second faint structure that propagates outwards. 

%If the eruption was more on limb in STEREO-B than in STEREO-A, the activity observed in AR 11029 could be related to the CME. Is it possible that the estimated source region incorrect? 

%Photospheric evolution 
Finally, we looked into the evolution of the photosphere. The red cross in Figure \ref{fig:eit} indicates the source of the stealth CME as estimated by \citet{kilpua2014solar}, whilst the blue crosses indicate the approximate footpoints of the CME found from the GCS model. The approximated source location is an area of weak dispersed field with no clear polarity inversion line, reinforcing the conclusion that the stealth CME originated from a nearby active region. Active region 11029 first starts to emerge early on 22 October 2009 in the eastern hemisphere, into a region of weak mixed polarity field. The active region builds up as the result of the emergence of several bipoles that coalesce. A second bipole begins to emerge $\sim$03:00 UT 24-Oct-2009 on the north-western side of the first bipole. A third bipole begins its emergence $\sim$10:00 UT 26-Oct-2009 at the polarity inversion line of the second bipole. Flux emergence is still underway at the time of the stealth CME as determined using the EUV data. At the time of the eruption the active region has a beta-gamma configuration according to the Hale classification scheme \citep{hale1919}, meaning that the region was bipolar overall but that no continuous line could be drawn separating spots of opposite polarities.

%Here we discuss the magnetic field configuration and evolution of active region 11029, the region we propose to the the source.
%A newly emerged active region is located west of the source region, whilst a decaying active region is located east of the source region, these two active regions are the source regions of numerous eruptions in the lead up to the stealth CME. 

\subsubsection{Overall Remarks}
%Suggest that the stealth CME is either from the active region, or triggered by the activity in the active region. 

The combined analysis of the photospheric magnetic field, activity in the lower corona and CME propagation as seen in coronagraph data together suggest that the stealth CME of 27 October 2009 originated in active region 11029. The CME is observed to be in its slow rise phase as observed by STEREO-B COR1 data by 23:00 UT 26-Oct-2009. The fast rise phase is observed to start around 13:00 UT 27-Oct-2009, similar in time to the formation of new loops in the active region, which are deemed to be post-eruption loops. The time of the fast rise phase onset is $\sim$7 hours later than the estimated time of eruption determined by \citep{kilpua2014solar}. This discrepancy may be partly due to the stealth CME originating from a high altitude, as determined from the absence of any signature of the erupting CME in EUV data, and only the underside of the CME being observable in COR1 data.

%  no sign of a propagating structure in the EUVI height-time profiles. No PIL in estimated source region - can a CME have been initiated from here?

  \begin{figure*}[t!]
\centering
    \includegraphics[width=1\linewidth]{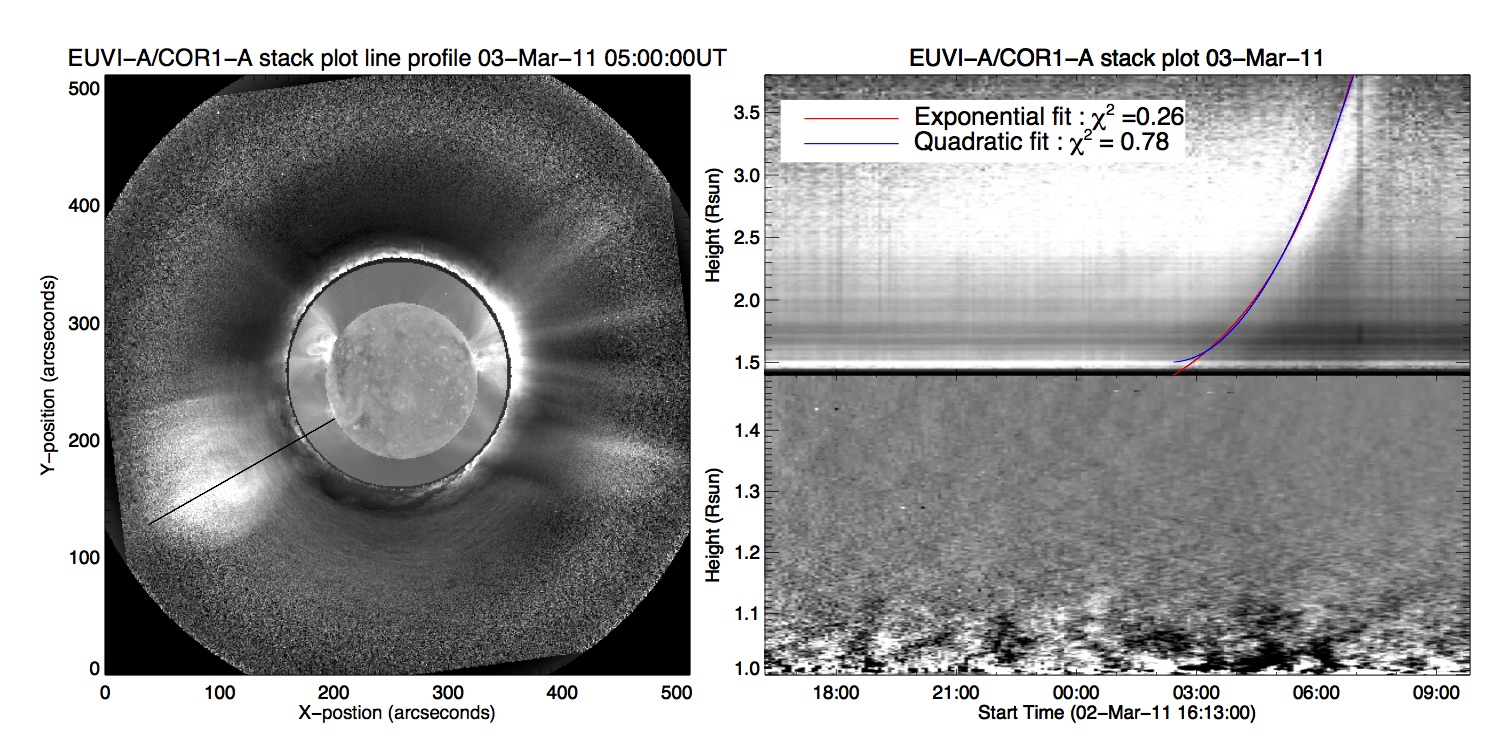}
    \caption{(Left) EUVI/COR1 image indicating the slice used to create the stack plot.  (Right) EUVI-A 195\AA (lower) and COR1-A (upper) stack plots.  The height-time profile of the stealth CME observed on 3rd March 2011 can be determined from the COR1-A data which show the cavity underside (concave-up structure) of the stealth CME. Exponential (red line) and quadratic (blue line) fits have been applied to the COR1-A data.}
    \label{fig:ht_se1}
  \end{figure*}
  
\begin{figure*}[t!]
\centering
    \includegraphics[width=1\linewidth]{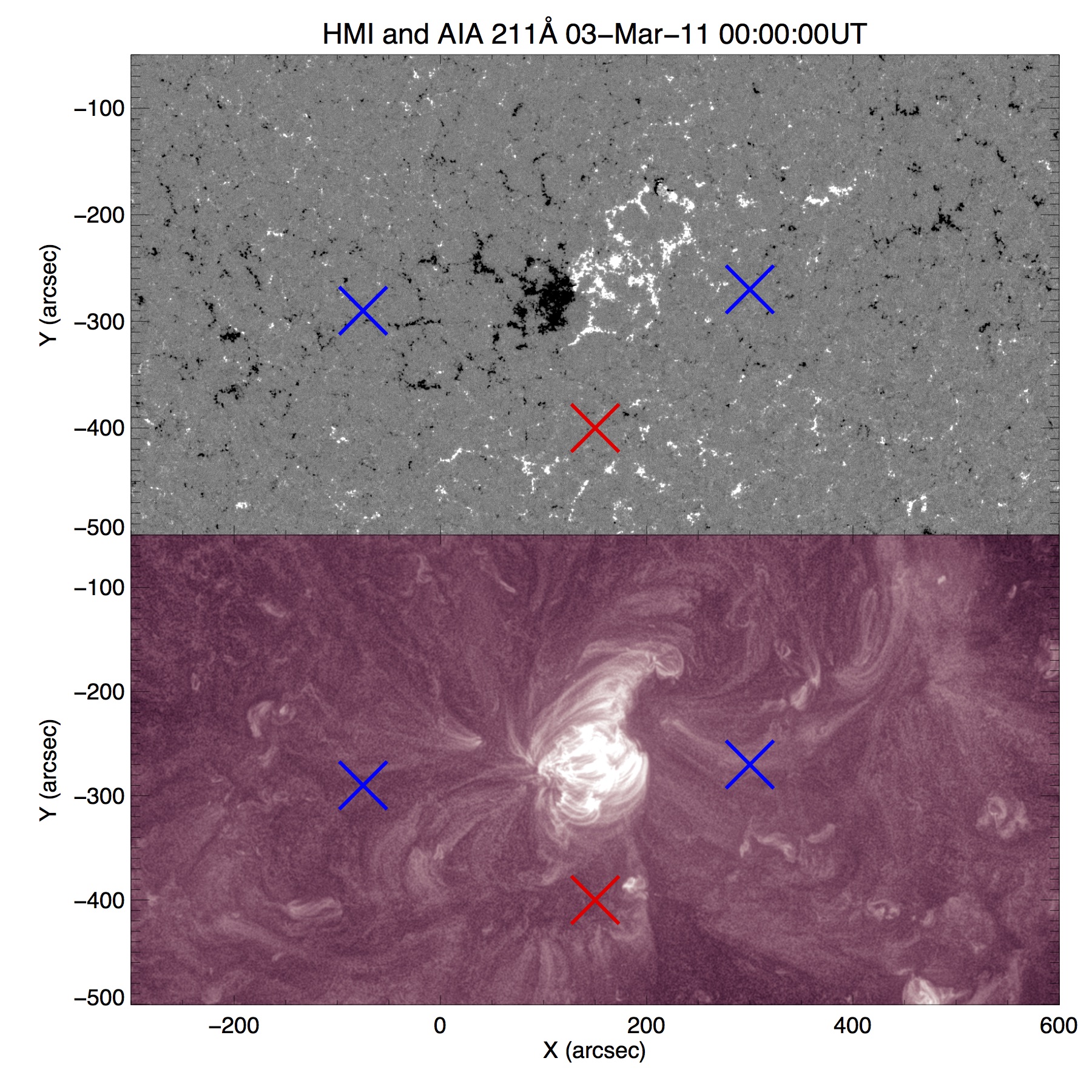}
    \caption{Top panel: HMI Magnetograph. A small active region (NOAA 11165) is located in the center of the image, with a polarity inversion line in a north-south direction. A second polarity inversion line runs in a east-west direction, where a filament channel is present. Bottom panel: MGN-processed SDO/AIA 211\AA\ image. Active region NOAA 11165 is located in the center of the image. The filament channel runs in a east-west direction, south-east of the active region. A filament lies to the west of the active region. The red X represents the triangulated region \citep{pevtsov2011coronal}. The two blue X's represent the footpoints of the structure as derived from the GCS model}
    \label{fig:aia}
  \end{figure*}

\subsection{Stealth event 2: 03-March-2011}  

\begin{figure*}[t!]
\centering
    \includegraphics[width=1\linewidth]{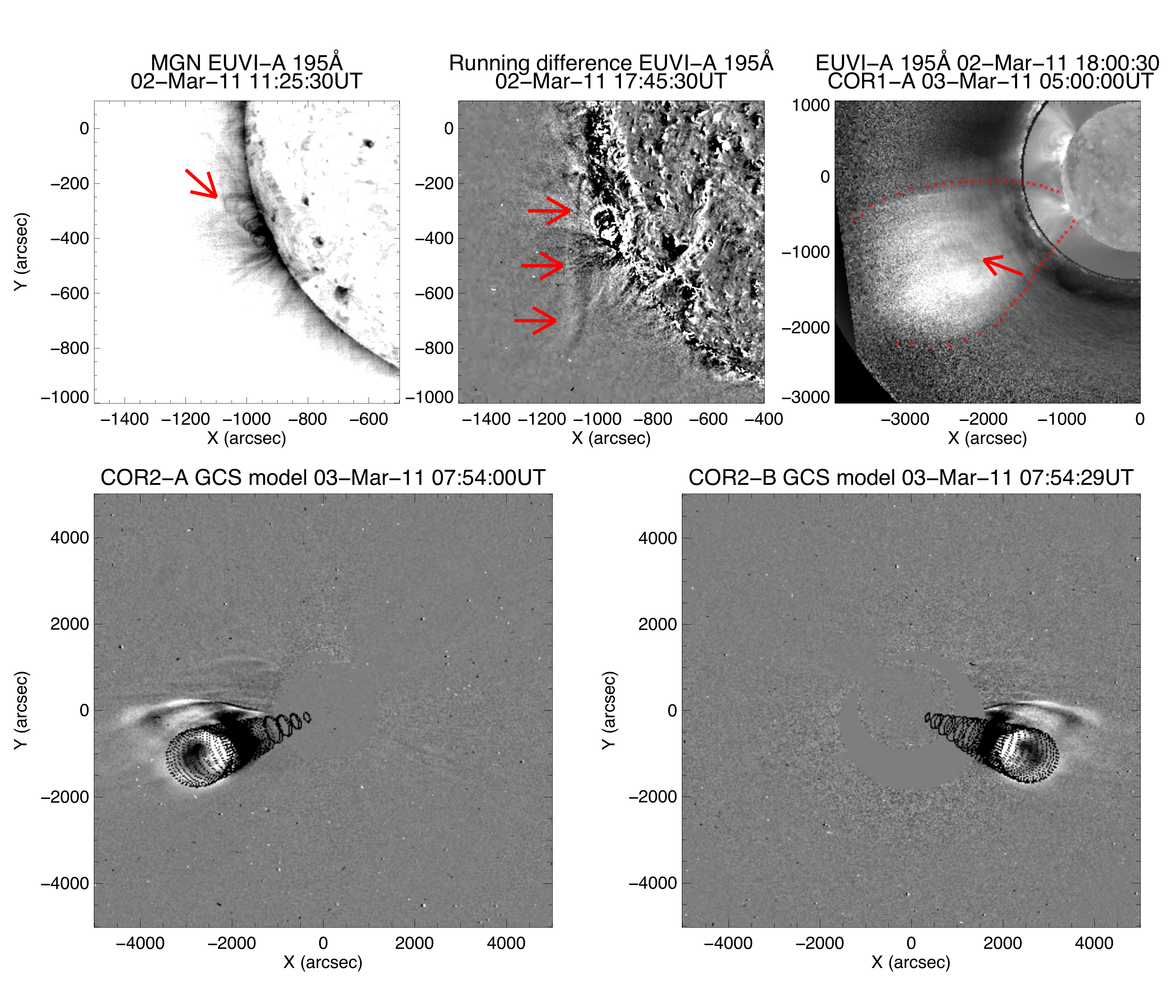}
\caption{Top : (Left) MGN processed EUVI-A 195\AA\ image. An expansion of bright loops were observed to the north of the active region, indicated by the arrow. (Center) Time difference EUVI-A 195\AA\ image, with a 30-minute temporal separation. A structure was observed stretching from the north to the south, and traveling outwards from the solar disk, \corr{indicated by the red arrows.} (Right) Combined EUVI-A 195\AA\ and COR1-A image. The combination illustrates how the bulb-structure in the EUVI images prior to the eruption expands into the CME structure observed in the COR1 images, outlined by the two dotted lines.  Bottom : The wire frame fitted to the CME in COR2-A (Left) and COR2- B (Right).}
\label{fig:stereo}
\end{figure*}

On March 3rd 2011 the STEREO coronagraphs observed an Earth directed CME with a faint leading edge that appeared to be slowly rising in STEREO-A COR1 data from around 00:00 UT. Despite the faint leading edge, the CME is seen to have a clear circular shape, the concave-up section of which was first observed in STEREO-A COR1 at approximately 03:00 UT on 03-Mar-2011. The CME had a plane-of-sky speed of 409 kms$^{-1}$ as seen by STEREO-A\footnote{http://solar.jhuapl.edu/Data-Products/COR-CME-Catalog.php}. The LASCO coronagraphs observed this CME as a faint partial halo that propagated to the south. This stealth CME was previously studied by \citet{pevtsov2011coronal} who found, using a triangulation approach, the CME source region to be S35W10. In the vicinity of the approximated source region was a small active region (NOAA active region 11165) and a filament channel as can be seen in Figure \ref{fig:aia}. \citet{pevtsov2011coronal} concluded that the filament channel was the CME source region. \citet{nitta2017earth} find two EUV dimmings centered around S20, the region in which the active region was present. The source region was on disk from the SDO perspective and at the solar limb from the perspective of both STEREO spacecraft. This means that the combined AIA and EUVI data allow the approximated source region to be studied when viewed at the limb as well \corr{as from above}. However, as detailed below, the STEREO-A EUVI data show more clearly the evolution and eruption of the source region and are focused on in this study.  
%In this section we present observations that discuss why we propose NOAA active region 11165 to be the stealth CME source region.

%, which erupted 23 hours after the eruption of a filament adjacent to the active region, on its western side. EUVI data prior to the stealth CME show emission exhibiting a bulb-shape structure was present at the same latitude as active region 11165. This structure gradually moved outwards over several hours and was no longer present after the eruption.

%From initial inspection of EUV images using AIA/SDO and EUVI/STEREO data, no eruptive signatures were observed, however on closer inspection of EUVI images prior to the eruption, emission exhibiting a bulb-shape structure was present at the same latitude of the small active region. This structure gradually moved outwards over several hours and was no longer present after the eruption. Additionally, 23 hours prior to the stealth CME, a filament eruption occurred to the west of the active region. 
  
\subsubsection{Analysis}

%Stack plot
%REF EDIT: again trying to address the initiation from high altitude. In previous event, the slow rise phase being captured in COR1 further led us to believe the initiation was a high altitude, however at this stage in the paper for event2 we only have the lack of structure in EUVI suggesting that - but later when we look at the radio date, the low frequencies will further suggest a high altitude structure for event2, and the text will be edited accordingly to say that
The stack plot shown in Figure \ref{fig:ht_se1} used a slice of the EUVI data and COR1 located at an angle of 240$^\circ$ clockwise from solar north (indicated by the black line in the left panel). The propagation of the concave-up section of the CME is clearly visible in the COR1-A stack plot (top panel of Figure \ref{fig:ht_se1}) but shows that the slow rise to fast rise transition was not captured. The slow rise to fast rise transition of the underside of the CME was not visible in the EUVI data, presumably because of insufficient plasma emission. This, combined with little to no structure observed in the EUVI, \corr{may be the result of} the CME being initiated from a high altitude structure with weak plasma emission. Although the CME propagation profile was fitted with an exponential and a quadratic curve, neither curve fits the CME curve better than the other, and therefore no conclusions on the most likely initiation mechanism can be made from this. 

\begin{figure*}[t!]
\centering
    \includegraphics[width=1\linewidth]{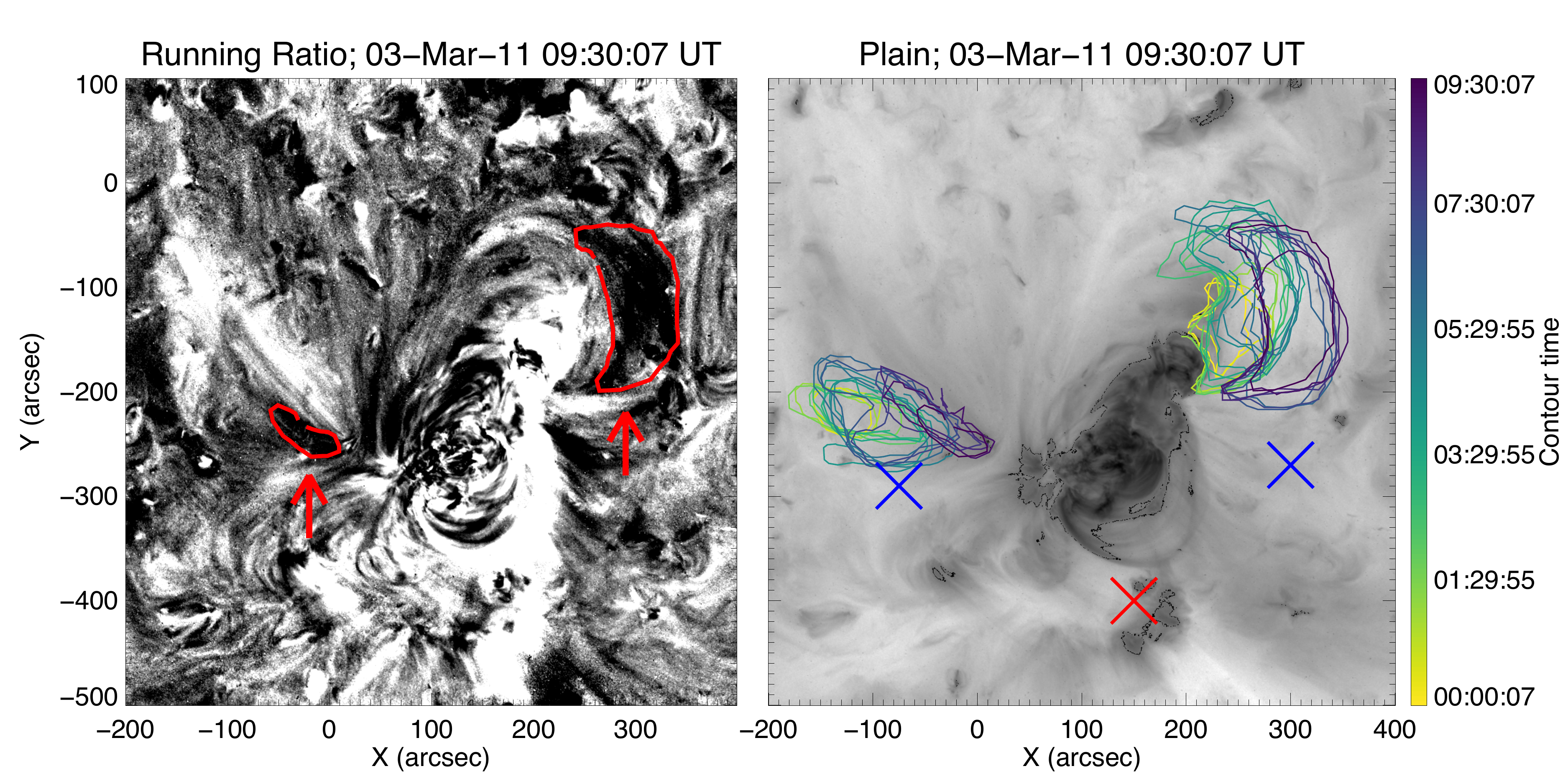}
    \caption{(Left) AIA 211\AA\ Running ratio image at the final time of EUV dimming tracking. (Right) AIA 211\AA\ plain image with inverted colour table showing the evolution of the dimming regions north and east of the active region, outlined with contours. The dimming regions evolved over a 9 hour period. Both images are de-rotated to the start time; 00:00 UT 03-Mar-2011}
    \label{fig:dim_n}
  \end{figure*}

%Coronagraph Observations?

%GCS MODEL
The GCS model was applied in the same way as discussed previously (Figure \ref{fig:stereo}). The footpoints of the eruption from the GCS model were found to be S18E04 and S16W18, illustrated by the blue X's in Figure \ref{fig:aia}. This was to the East and West of the small active region (NOAA 11165), and further North than the region triangulated by \citet{pevtsov2011coronal}. The footpoints are of a similar location to the dimmings found by \citep{nitta2017earth} centered around S20.

%EUV observations
The MGN-processed AIA 211~\AA\ images show dynamic plasma emission structures that occur around active region 11165. Likewise, MGN-processed images and running difference images with a 30 minute temporal separation from EUVI-A 195~\AA\ show a number of dynamic structures located off-limb and out to the edge of the field of view above active region 11165. Comparing 195~\AA\ EUVI-A and AIA 211~\AA\ observations, we have been able to identify various dynamic structures that temporally and spatially correlate between the data sets and which together enable an investigation of the corona in the lead up to the time of the eruption. Around 17:00 UT on 01-Mar-2011, a filament is observed to begin to rise and gradually erupt over approximately a 6-hour period. It is noted that the CME produced by this filament eruption is observed in the COR1-A field of view approximately 23 hours before the stealth CME. The filament eruption creates new loops that connect the eastern side of the filament channel with the west side of active region 11165.
%EUV loops
The AIA data show that from $\sim$05:00 UT on 02-Mar-2011, a number of loops at the periphery of the active region 11165 begin to reconfigure. On the east side of the active region, a loop is observed to have been disconnected from an area in the northern part of the active region. It then swings up and over the active region in an anti-clockwise direction at $\sim$05:15 on UT 02-Mar-2011. This structure is seen in EUVI data to be almost parallel to the solar limb and in motion at $\sim$06:10 UT on 02-Mar-2011. The loop expands with a north-south motion and is shortly followed by the creation of a new, larger-scale loop system in the north of the active region. These new loops are observed in both the AIA and EUVI-A data (Figure \ref{fig:stereo}, left-hand panel). 
%Flare ribbons and flare loops
The activity observed in active region 11165 using EUV data also includes the formation of a pair of faint flare ribbons and their associated loops, that are located at the edge of the magnetic bipole away from the internal polarity inversion line. AIA 211~\AA\ and AIA 304~\AA\ data indicate that the flare ribbons form at $\sim$09:00 UT on 02-Mar-2011, with a second phase of brightening and expansion away from the centre of the active region at $\sim$21:40 UT on 02-Mar-2011. This location and evolution indicates the occurrence of magnetic reconnection in a region above the active region loops.
%plasma flow
A flow of plasma is seen moving out from the south of active region 11165 from $\sim$08:35 UT on 02-Mar-2011, and a second flow follows at $\sim$13:55 UT on 02-Mar-2011. The visible end of the second flow appears to be immediately followed by a structure that is again almost parallel to the limb, stretching across from the north to the south, and expanding outwards from $\sim$17:50 UT on 02-Mar-2011 (Figure \ref{fig:stereo}, middle panel). Fainter flows consequently move outwards for a short period from $\sim$22:10 UT on 02-Mar-2011. 
%Post-eruption loops
New loops slowly begin to form in the early hours of 03-Mar-2011 and, given their close association with the CME observed in STEREO-A COR1 data and EUV dimmings, they are termed post-eruption loops. These post-eruption loops grow larger over a 6-hour period. Combined EUVI-A 195~\AA\ images and COR1-A images at different times show that a bulb-shaped structure with a roughly circular centre is observed to move out of the EUVI-A images into COR1-A images where it becomes the stealth CME studied here (Figure \ref{fig:stereo}, right-hand panel). The clear circular cavity present in the center of the CME seen in the COR1-A is suggestive of a flux rope configuration at this time.

\begin{figure*}[t!]%
    \centering
    {{\includegraphics[width=0.47\linewidth,trim=35 15 130 75,clip]{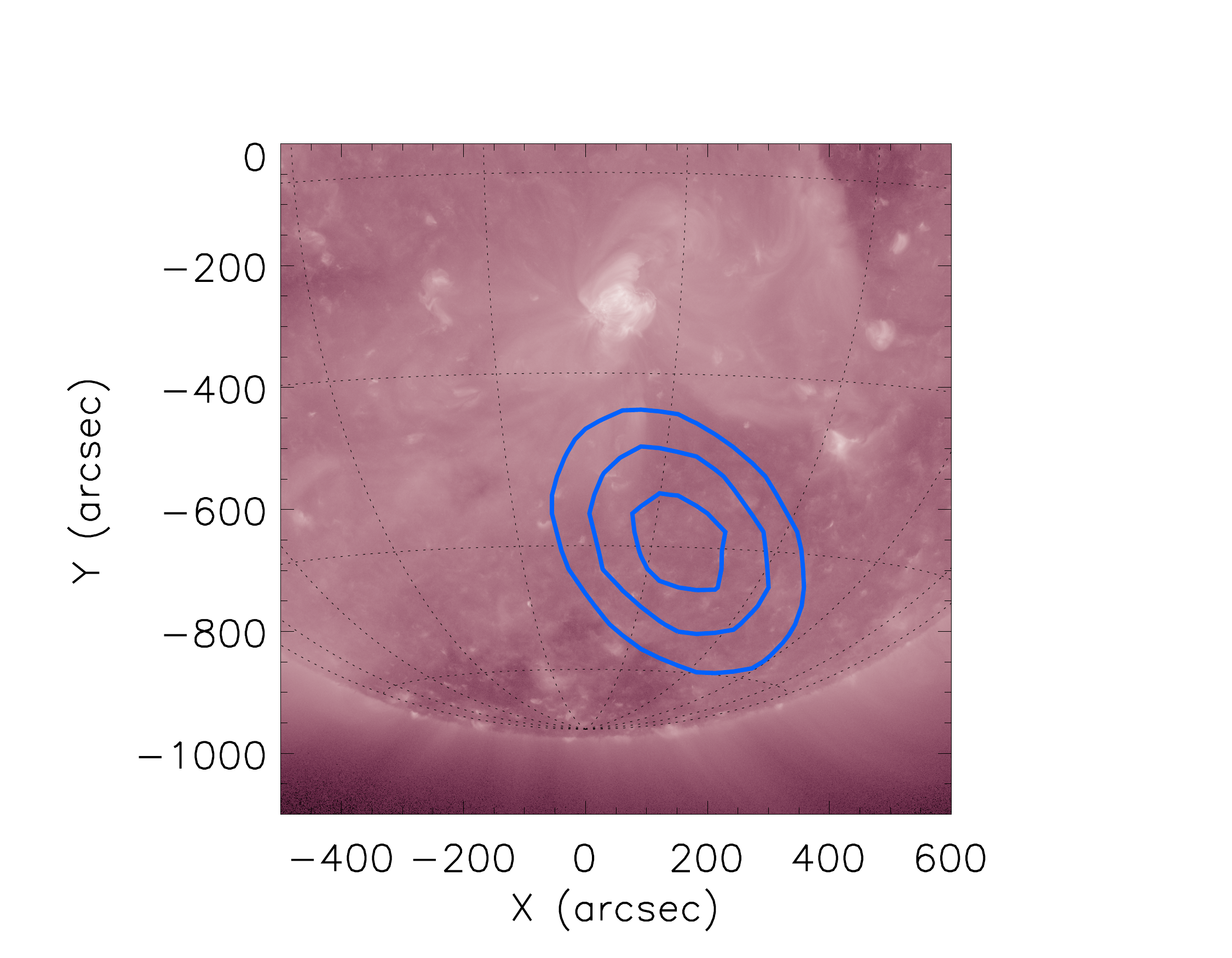}}}%
    \qquad
    {{\includegraphics[width=0.49\linewidth,trim=65 15 35 35,clip]{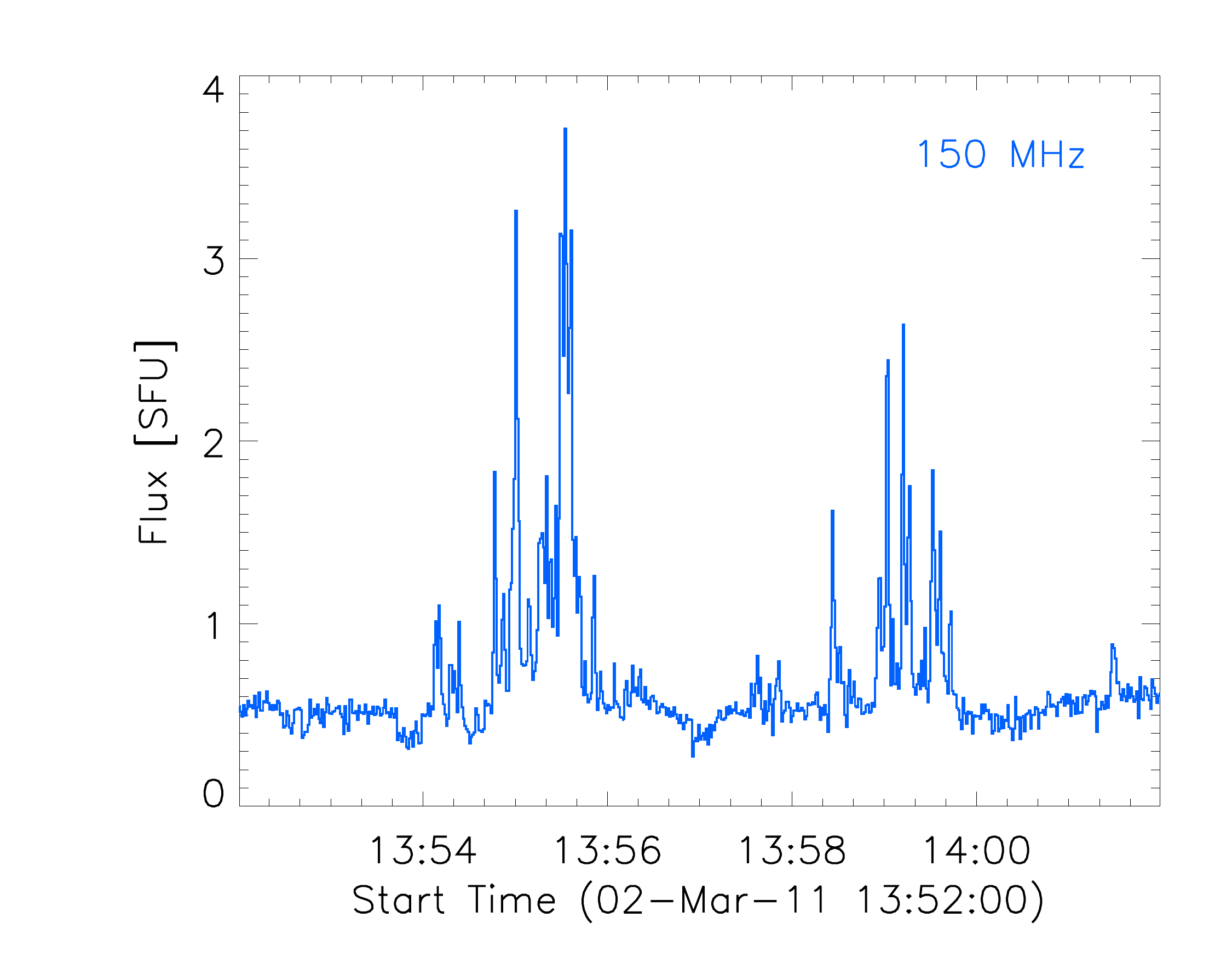} }}%
    \caption{Left : Image at 13:55 UT showing the 50, 70 and 90 percent 150 MHz radio contours superimposed on the AIA 211~\AA\ image. Right : The flux profile at 150 MHz}%
    \label{fig:radio}%
\end{figure*}
 
%\subsubsection{Evolution of twin EUV dimmings}
%\label{subsebsec:EUVdimmings_mar2011}

Running difference images and running ratio images with 30 minute temporal separation revealed two faint dimming regions located either side of active region 11165 (Figure \ref{fig:dim_n}a), which both match with those found by \citet{nitta2017earth}, and are not too far from where the footpoints estimated from the GCS model were found to be. We can therefore conclude that these two dimmings are representative of the footpoints of the erupting stealth CME. As can be seen from Figure \ref{fig:dim_n}b, the dimming region to the north-west of the active region both grew in size and moved away from the active region between 00:00 UT and 09:30 UT 03-Mar-2011, whilst the dimming region to the east of the active region grew in size initially, and then shrunk back down in size. There is an indication that the dimmings underwent a slight clockwise rotation between 00:00 UT on 03-Mar-2011 and 09:30 UT on 3-Mar-2011. It is notable that the eastern dimming region is located in the same area as the footpoint of the dynamical structure that pulled and twisted out at $\sim$05:15 UT on 02-Mar-2011, indicating a connection between the structure involved in the dynamics prior to the CME and the erupting field. 

%The dynamic structures observed in the EUVI images therefore may be playing more of a roll in the formation of a high altitude structure? 
%Photospheric evolution 
Looking at the photospheric evolution of the region, NOAA active region 11165 began to emerge on the Sun on 25 February 2011 in the eastern hemisphere and into the magnetic field of a previously decayed active region. Active region 11165 emerged at the polarity inversion line of the decayed pre-existing region and with the same field orientation (positive leading magnetic field). At the time of the stealth CME on 3 March 2011, the active region had a bipolar configuration and very dispersed magnetic field having been acted on by super-granulation. Two episodes of flux emergence occurred in active region 11165 between its first appearance on disk and the time of the stealth CME, at 22:40 UT on 25-Feb-2011 and 06:30 UT on 28-Feb-2011. The evolution of the photospheric field is dominated by flux emergence rather than flux cancellation. The polarity inversion line above which the CME originated (as determined from dimmings and post-eruption loops) was oriented in a north-south direction, indicating that differential rotation had not yet had a significant effect on the active region's configuration (Figure \ref{fig:aia}). To the south of the active region lies a polarity inversion line that is associated with an (empty \citep{pevtsov2011coronal}) filament channel. This inversion line was initially \corr{thought to be the} location of the origin of the stealth using triangulation \citep{pevtsov2011coronal}. 

%The region was tracked back to when it appeared from around the limb of the Sun. From this point, it is clear to see that the area is full of highly dispersed weak field and the small active region is not present at this point. A small amount of flux begins to emerge into the decayed field, this emergence continues and the polarities separate out. The active region is formed by the -Feb-2011. Flux continues to emerge into the active region in the lead up to the eruption. 

%Radio data from hamish
%REF EDIT : the radio emission suggests high alt structure - backing up suggested conclusion from stack plots
Lastly, we looked at the radio emission of the region. The second flow observed in EUV data at $\sim$13:55 UT on 02-Mar-2011 coincides with a brightening in radio frequencies around 150 MHz. The brightening was imaged by the Nan\c{c}ay Radioheliograph \citep{1997LNP...483..192K} most prominently at 150 MHz between 13:52 and 14:02 UT (Figure \ref{fig:radio}).  The emission arises from a source that appears to the south of active region NOAA 11165, as viewed in the plane of the sky. This spatially and temporally corresponds to the second flow observed in EUV data. The impulsive nature of the radio emission implies that particle acceleration occurs in conjunction with this second flow of plasma that is seen around 13:55~UT.  Assuming second harmonic plasma emission, as the polarisation is less than 10\%, the 150 MHz emission corresponds to an altitude of 0.34 solar radii (238 Mm) using the Newkirk coronal density model \citep{1961ApJ...133..983N}.  The emission is not observed above 173 MHz by the Nan\c{c}ay Radioheliograph, restricting it to these higher coronal altitudes. This \corr{coincides with the lack of structure observed in the EUVI stack plots being a result of a high altitude structure}, supporting a hypothesis that the stealth CME was ultimately the result of a reconfiguration of high-altitude magnetic field, above the active region core, that may have involved magnetic reconnection. The faint radio emission is not visible at lower wavelengths below 150 MHz, detected by full-sun spectrometers, so we cannot confirm whether the emission is a type III burst, caused by propagating electron beams (e.g. \citet{2014RAA....14..773R}), or localised electron acceleration more in-line with a type I burst \citep{1985srph.book..415K}.

\corr{Normal CMEs can have a multitude of accompanying radio emission, particularly from the upper solar corona.  The stealth CMEs do not have any accompanying type II radio emission.  The slow speed of stealth CMEs means that we do not expect it to drive a shock, where shock-driven acceleration can generate type II radio emission.  Faster, more intense CMEs can also display moving type IV emission \citep{james2017disc}, generated via gyrosynchrotron emission by trapped, high-energy particles within the CME.  Normal CMEs that have associated flares are frequently accompanied by type III bursts, signatures of accelerated electron beams escaping the Sun.  Given the apparent high altitude of the stealth CMEs, if any electron beams are accelerated during the magnetic instability that initiates the CME liftoff, we might expect to detect faint, lower frequency ($<$100 MHz) type III emission.  The emission that we observe on March 3rd is very \dorr{faint}, and brief considering the duration of the stealth CME lift-off, with no low-frequency emission observed using the full-disc integrated radio spectrometers.  A future imaging spectroscopy task for the new, high sensitivity radio interferometers like the Low Frequency Array (LOFAR) \citep{van2013lofar} and the upcoming Square Kilometre Array (SKA).}

\subsubsection{Overall Remarks}

The combined analysis of the activity in the lower corona as observed on disk and at the limb, and the CME propagation as seen in coronagraph data together indicate that the stealth CME of 3 March 2011 originated in active region 11165. The MGN technique enhanced subtle changes in the evolution of plasma emissions structures consistent with changes in the magnetic field configuration of the active region. EUV dimmings at the periphery of the active region, the lack of opening of the active region arcade field and the side-on view afforded by the STEREO spacecraft reveal that the erupting structure originated at a relatively high altitude above the core active region loops that dominated the EUV emission. The lack of significant flux cancellation in the active region also suggests that the scenario of low altitude flux rope formation and eruption of \citet{van1989formation} does not occur here.

% Due to the dynamical features observed in the processed images, the faint EUV dimmings either side of the small active region, and post eruptive loops over the active region. No clear structures in the height-time plots, is the structure forming at a high altitude? being Initiated from a high altitude? No flux cancellation observed so either a flux rope was formed early on or is being formed at a higher altitude. No major flaring, reconnection potentially slower and at a higher altitude in the corona where it is less dense, less particles to accelerate? Triggered either by flux emergence of the new active region, or by the filament eruption. But the filament eruption may happen too early for it to account for the stealth CME. 

%\startlongtable
\begin{deluxetable*}{lcc}
\centering
\tablecaption{Table summarising the timeline of the activity evolution related to the two stealth CME events.}
\tablehead{                   & Event 1: 27 October 2009            & Event 2: 3 March 2011}
\startdata
 Previous adjacent eruption   & 11:25 UT 26 October 2009            & 17:00 UT 1 March 2011  \\
 Flares and/or ribbons        &                                     & 09:00 UT 2 March 2011   \\
                              &                                     & 21:40 UT 2 March 2011      \\
 Slow rise phase              & $\sim$23:00 UT 26 October 2009      &         \\
 Flares and/or ribbons        & $\sim$04:49 UT 27 October 2009      & 09:00 UT 2 March 2011   \\
                              & $\sim$07:05 UT 27 October 2009      & 21:40 UT 2 March 2011     \\
 Fast rise phase onset        & 13:00 UT  27 October 2009           &           --            \\
 Dimming onset      		  &          --                         & 00:00 UT 3 March 2011   \\
 Post-eruption loop formation onset &  13:00 UT 27 October 2009  & 00:00 UT 3 March 2011   \\
\enddata

\label{table:summary}
\end{deluxetable*}

%TESTING FIGURES
\begin{figure*}[t!]%
    \centering
    {{\includegraphics[width=1\linewidth]{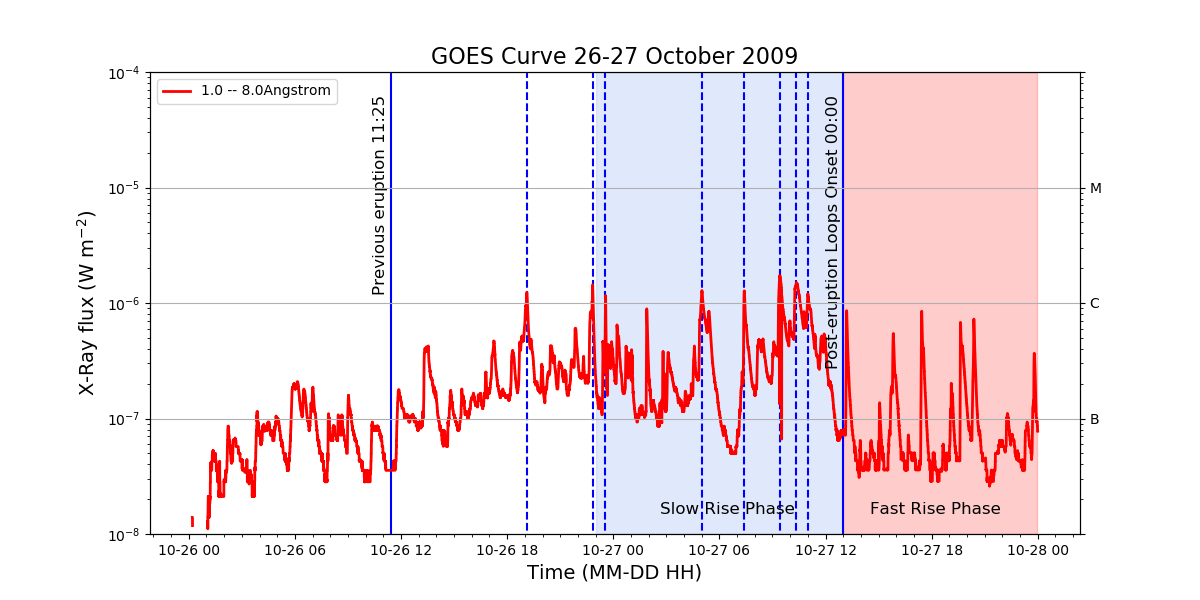} }}%
    \qquad
    {{\includegraphics[width=1\linewidth]{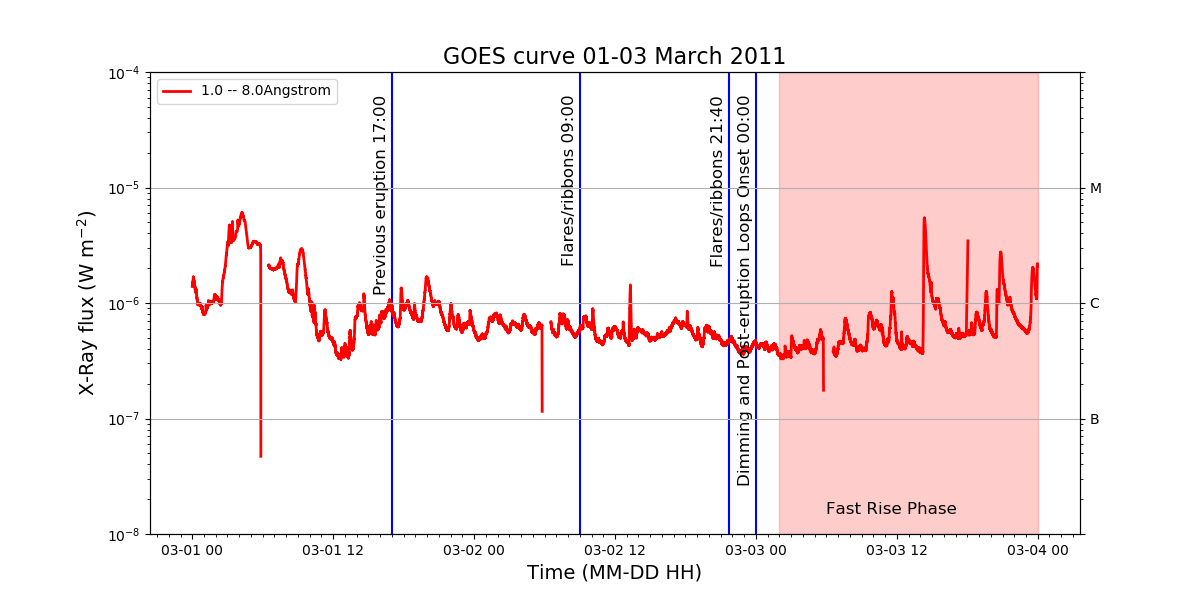} }}%
    \caption{Top : Annotated GOES curve for the time leading up to the stealth CME of 27th October 2009. Active region 11029 was the only active region on disk at this time. The dashed lines represent C-class flares from AR 11029. The light blue shaded region represents the time of the slow rise phase of the stealth CME and the light red shaded region represents the CME fast rise phase. Bottom : Annotated GOES curve for the time leading up to the stealth CME of 3rd March 2011. There were multiple active regions in the northern hemisphere during this time, which contribute to the GOES light curve. The light red shaded region represents the stealth CME fast rise phase, the slow rise phase was not captured.}%
    \label{fig:GOES}%
\end{figure*}

In the hours leading up to the stealth CME, active region 11165 and the surrounding corona undergo some notable activity. First is the eruption of a filament immediately adjacent to the active region on its western side. The filament eruption reconfigures the coronal field and produces new connections between the filament channel and active region 11165. Such an eruption may have altered the corona above the active region. In addition, prior to the stealth CME, activity was also in active region 11165 itself. The formation of new EUV loops from $\sim$05:00 to $\sim$06:00 UT on 02-Mar-2011, without any observed flux emergence at this time, are suggestive of coronal reconnection, which may have played a role in creating the pre-eruptive field. 

\section{Discussion} \label{sec:discussion}

% Summary and MGN technique successfully showed lower coronal signatures
Both events were determined to be Earth directed using triangulation and graduated cylindrical shell modelling. MGN image processing was applied to SOHO/EIT, SDO/AIA, and STEREO/EUVI data.  As shown by \cite{alzate2017identification}, such an approach revealed subtle changes in the coronal emission structures that were not identifiable in the unprocessed data. For example, within the EUVI-A processed images, several structures were seen moving outwards from the solar limb, the timings of which correlated to dynamic structures observed in the sharpened AIA images as seen on disk and the white light CMEs seen in the coronagraph data. Indeed, the enhanced EUV images were able to reveal several observational phenomena that are in line with the CSHKP standard model of an eruption, including dimming regions and post-eruption arcades. Such lower coronal signatures were observed in both stealth CME events and indicate that they originated in NOAA active regions 11029 and 11165, respectively, contrary to previous work that has suggested stealth CMEs might in some way be associated with open magnetic field regions, the quiet sun, or empty filament channels. Our findings support previous work by \cite{alzate2017identification} who showed imaging processing techniques are essential in searching for the origins of CMEs that leave no obvious signatures on disk. 

%The occurrence of CSHKP signatures also reinforces the suggestion that the basic mechanisms involved in non-stealth CMEs are taking place in stealth CMEs. Particularly the role of magnetic reconnection in in the eruption process that transforms a sheared arcade into a flux rope, which adds poloidal flux to the erupting structure and accelerates the CME away from the Sun. EXPAND AND MORE CITATIONS NEEDED.

%stealth CME models: what do our observations tell us about how stealth CME obs hold up
% against models?
 The identification of low coronal signatures associated with the CSHKP model enables us to not only identify the source region of stealth CMEs, but also to analyse the evolution of each region in the time leading up to the eruption for comparison with CME theories. Models that have specifically been suggested as being relevant to stealth CMEs are those of streamer blowout CMEs \citep{howard2013stealth} that invoke differential rotation as the mechanism that energises the magnetic field system to bring it to an eruptive state \citep{vourlidas2018streamer}. In this scenario, the pre-eruptive magnetic field configuration could be that of an arcade or a flux rope. In the streamer blowout numerical model of \citet{lynch2016model}, the pre-eruptive magnetic field configuration is that of a sheared arcade energised via photospheric shearing motions within a multipolar field configuration. The shearing motions lead to breakout reconnection above the central arcade followed by flare reconnection within the sheared arcade, which forms a flux rope and accelerates the CME. The key role of photospheric flows in the above mentioned model implies that extended polarity inversions lines should be present in stealth CME regions. We note that we do not see such extended size scales in either of our stealth CMEs. 
 %An estimate of the source region of each stealth CME in this study does show that the event of Oct-2009 is very mixed field with no clear PIL. Mar-2011 event, approximated source region has a very extended PIL. 
In light of this we will go on to analyse further the source region characteristics and discuss them in the context of CME models in general.

% Our stealth CMEs are not from extended PILS, but from extended heights
 Even though both stealth CMEs originated in active regions without extended polarity inversion lines, they were both formed in a magnetic field configuration that was extended in altitude, as found by \cite{robbrecht2009no,d2014observational,alzate2017identification}. For example, STEREO data for the 27 October 2009 event showed that the underside of the erupting structure (as determined from the concave-up feature) was clearly visible at 0.5R$_{\odot}$ above the photosphere during its slow rise phase, with the transition to the fast rise phase occurring at $\sim$1R$_{\odot}$. These values are similar to those found in \cite{robbrecht2009no}. Such a high altitude means that the erupting structure originated in a region with lower plasma density and weaker magnetic field than is usually found for active region CMEs. The eruptions did not originate in the core active region magnetic field of NOAA active regions 11029 and 11165 that is responsible for the dominant active region EUV or soft X-ray emission. This high altitude location in turn leads to CMEs that have low accelerations due to the low magnetic field strength. 

%Height-time evolution of erupting structures
% coronagraph data: cavity and flux rope interpretation
% flux rope structure during the slow-rise phase for 27 Oct. 2009 event
 For both events COR1-A stack plots show the propagation of the \textit{underside} of the stealth CMEs, not the leading edge of the erupting structure. However, it was not possible to identify the underside of each erupting structure in the EUVI-A stack plots, presumably due to weak plasma emission in the 195\AA\ waveband associated with a low plasma density due to their high-altitude. It is notable that when the stealth CME of 27 October 2009 is in its slow-rise phase (\corr{that is}, before the reconnection associated with the fast-rise phase sets in) a flux rope is already present. The flux rope is identified through the concave-up structure seen in STEREO coronagraph images and this observation is supportive of a pre-eruptive flux rope having formed. Both events show a clear cavity with a concave-up structure, indicating the presence of a flux rope, during the fast rise phase. This is expected since regardless of the pre-eruptive configuration, flare reconnection within a sheared arcade will always build a flux rope.

%when and how does the flux rope form in event 1?

 The challenge now is to try and discern whether any aspects of the evolution of NOAA active region 11029 that produced the 27 October 2009 event can be linked to the formation of the flux rope prior to its slow-rise phase. There is increasing observational support for the importance of the role of magnetic reconnection in the formation of eruptive structures. Observationally this is manifested  by (confined) flaring or flux cancellation that is able to transform a sheared arcade into a flux rope. The height of the reconnection then determines the height at which the underside of the flux rope is located from photosphere/chromosphere \citep{chintzoglou2015formation} into the corona (e.g. \citep{james2017disc,james2018}). SDO/HMI and SOHO/MDI images of both stealth CME source regions in this study showed no major or sustained flux cancellation in the time leading up to eruption, nor were any S-shaped plasma emission structures observed in the the AIA, EIT or EUVI passbands that may have indicated the formation of a flux rope that could then have risen above the active region core. However, NOAA active region 11029 showed weak flaring in the hours leading up to the stealth CME (table 2, top panel figure \ref{fig:GOES}). The flaring will have been due to reconnection in the corona and this reconnection could have produced the flux rope  in the corona. Even though there is no observational support for the presence of a pre-eruptive flux rope for the stealth CME event of 3rd March 2011, it is notable that flaring and flare ribbons are observed in NOAA active region 11165 in the hours leading up to the eruption. Both stealth CME source regions show a\corr{ similar} evolution in this regard.

%Although the flaring observed in the stealth CME source regions is indicative of such  a process, we have insufficient observations to support this in our events. Since we cannot track the slow-rise phase of the march 2011 stealth CME, we are not able to say when the flux rope formed. It may have formed during the eruption itself similar to the flux rope observed by \citet{robbrecht2009no} and modelled by \citet{lynch2016model}.

%Sympathetic events
 Previous studies of stealth CMEs have suggested that they may be sympathetic eruptions triggered by a reconfiguration of overlying field, and therefore removal of stabilising flux, due to a nearby CME. Indeed, the 27 October 2009 event had an eruption from a nearby region 11.5 hours before its slow-rise phase was observed, and the event of 3 March 2011 had an adjacent eruption 31 hours before its CME-related dimming was observed (table 2, bottom panel figure \ref{fig:GOES}). In the absence of extended polarity inversion lines and sustained shearing motions, a sympathetic eruption suggests the presence of a pre-existing flux rope in quasi-equilibrium \citep{torok2011}, which is what we find for the event of 27 October 2009.

%energy storage and release process
 
 A fundamental aspect of CMEs is that they are known to be the result of an energy storage and release process \citep[see][for an overview]{green2018origin}. As discussed above, theoretical and modelling work on stealth CMEs has proposed that the energy injection is provided by slow shearing motions created by photospheric differential rotation. However, in this study the stealth CMEs come from active regions with a short polarity inversion line oriented in the north-south direction, and therefore not significantly acted on by differential rotation. However, the active regions in which both events originated exhibited flux emergence, which could have been the process by which the energy was injected. 

%New challenges
The question raised by this study is then whether
stealth events represent the high-altitude part of a spectrum of CMEs, related to flux rope formation by high-altitude magnetic reconnection. Structural changes occur above the core field of both active regions, and we suggest that the stealth CMEs originate from magnetic field whose footpoints are embedded either side of each active region. The vertical extent of each eruptive structure presents additional challenges in terms of reconstructing the magnetic field configuration from photospheric magnetic field extrapolations and numerical modelling to capture the formation of the structure. 

%Instrument limitations
The availability of 195~\AA\ data from both SOHO/EIT and STEREO/EUVI for the first stealth CME event means that it presents an interesting case study for investigating the possible role instrument capabilities play, as highlighted by \citet{howard2013stealth}. Despite applying the MGN image processing techniques to SOHO EUV data, little could be observed in comparison to the STEREO EUV data. Observational limitations are clearly shown in this case, as more features could be distinguished in the STEREO/EUVI data. Consideration should therefore be given to temperature response, dynamic range and image cadence during operation in the development of future EUV imagers. The key aspect is to design telescopes that are able to detect CMEs with weak signatures in EUV. This study supports the growing focus on the so-called middle corona and the need for instrumentation that can capture the evolution of structures with faint EUV emission. In the future, with more appropriate instrumentation, what would have been classed as a stealth CME in the SOHO or SDO era may no longer be the case. In addition, we find that the side-on view provided by STEREO was crucial in identifying, and studying the evolution of, the stealth CME source regions that were challenging to observe from above due to the dominance of emission from the active region core. Nonetheless, investigating what causes these events to have such weak signatures, but still produce magnetic structures that escape the Sun remains to be investigated further, and will aid overall understanding of the physical processes involved in CME initiation. 

\section{Conclusions} \label{sec:conclusion}

%Main finding
This study used advanced image processing techniques to identify and study the source regions of two stealth CMEs that were observed in multi-spacecraft coronagraph data on 27 October 2009 and 3 March 2011. We find that both stealth CMEs originated in active region areas as opposed to the quiet sun or in filament channels, contrary to \corr{the previous studies on these stealth events \citep{kilpua2014solar,pevtsov2011coronal}}. However, the erupting structures were not formed in the core active region field, but likely at altitudes of $\sim$0.5R$_{\odot}$ above the photosphere. The energy injection appears not to be the result of differential rotation but instead related to the emergence of new flux into the active region. In the event of 27 October 2009 we find observational support for the presence of a flux rope formed by reconnection in the corona during or before the slow rise phase of the CME. The flux rope may have been destabilised as a sympathetic eruption following a \dorr{nearby} CME. We find that the stealth CMEs of this study are no different to other CMEs in that they show features of the standard model but at the lower energy end of the spectrum with weaker signatures that current instrumentation can only just resolve. 

\acknowledgments

JRO thanks the STFC for support via funding given in her PHD studentship. JRO also thanks Mathew West and Marilena Mierla for useful discussions at the Royal Observatory Belgium.
LMG acknowledges support through a Royal Society University Research Fellowship and through the Leverhulme Trust Research P`roject Grant 2014-051. DML acknowledges support from the European Commission's H2020 Programme under the following Grant Agreements: GREST (no.~653982) and Pre-EST (no.~739500) as well as support from the Leverhulme Trust for an Early-Career Fellowship (ECF-2014-792) and is grateful to the Science Technology and Facilities Council for the award of an Ernest Rutherford Fellowship (ST/R003246/1). HASR acknowledges support from the STFC consolidated grant (ST/P000533/1). \corr{The authors thank the anonymous referee whose comments helped to improve the paper.}
SDO is a mission of NASA’s Living With a Star Program. STEREO is the third mission in NASA’s Solar Terrestrial Probes program. SOHO is a mission of international cooperation between ESA and NASA. The authors thank the SDO, STEREO, and SOHO teams for making their data publicly accessible.




%\newpage
\bibliographystyle{agsm}
\bibliography{Bib.bib}



\end{document}